\documentclass[twocolumn,trackchanges]{aastex63}
\usepackage{gensymb}
\usepackage{hyperref}
\usepackage{graphicx}

\turnoffedit

\accepted{PSJ}

\shorttitle{Europa's H$_2$O$_2$: Temperature Insensitivity and a Correlation with CO$_2$}
\shortauthors{Wu et al.}

\graphicspath{{./}{figures/}}

\begin{document}

\title{Europa's H$_2$O$_2$: Temperature Insensitivity and a Correlation with CO$_2$}

\correspondingauthor{Peiyu Wu}
\email{pw443@cornell.edu}

\author[0000-0002-8690-4910]{Peiyu Wu}
\affiliation{Department of Earth and Atmospheric Sciences, Cornell University, Ithaca, NY 14853, USA}

\author[0000-0002-0767-8901]{Samantha K. Trumbo}
\affiliation{Cornell Center for Astrophysics and Planetary Science, Cornell University, Ithaca, NY 14853, USA}
\affiliation{Department of Astronomy \& Astrophysics, University of California, San Diego, La Jolla, CA 92093, USA}

\author[0000-0002-8255-0545]{Michael E. Brown}
\affiliation{Division of Geological and Planetary Sciences, California Institute of Technology, Pasadena, CA 91125, USA}

\author[0000-0002-9068-3428]{Katherine de Kleer}
\affiliation{Division of Geological and Planetary Sciences, California Institute of Technology, Pasadena, CA 91125, USA}

\begin{abstract}
H$_2$O$_2$ is part of Europa's water-ice radiolytic cycle and a potential source of oxidants to Europa's subsurface ocean. However, factors controlling the concentration of this critical surface species remain unclear. Though laboratory experiments suggest that Europa's H$_2$O$_2$ should be concentrated in the coldest, most ice-rich regions toward the poles, Keck adaptive optics observations have shown the strongest H$_2$O$_2$ signatures in comparatively warm, salt-bearing terrain at low latitudes. As a result, it was suggested that the local non-ice composition of these terrains---particularly hypothesized enrichments of CO$_2$---may be a more dominant control on H$_2$O$_2$ than temperature or water-ice abundance. Here, we use observations of Europa from the NASA Infrared Telescope Facility, Keck Observatory, and JWST to disentangle the potential effects of temperature and composition. In order to isolate the effect of temperature on Europa's H$_2$O$_2$, we use the ground-based observations to assess its response to temperature changes over timescales associated with Europa's daily eclipse and diurnal cycle. We use JWST Cycle 1 data to look for any geographic correlation between Europa's H$_2$O$_2$ and CO$_2$. Both changes in Europa's 3.5-$\mu$m H$_2$O$_2$ absorption band from pre to post eclipse and across a local day suggest minimal effects of the local temperature on these timescales. In contrast, the JWST observations show a strong positive correlation between Europa's H$_2$O$_2$ and CO$_2$ bands, supporting the previously suggested possibility that the presence of CO$_2$ in the ice may enhance H$_2$O$_2$ concentrations via electron-scavenging.

\end{abstract}

\keywords{Galilean satellites (627), Planetary surfaces (2113), Surface composition (2115), Europa (2189), Infrared spectroscopy (2285)}

\section{Introduction} \label{sec:intro}
Hydrogen peroxide (H$_2$O$_2$) is a product of the important water-ice radiolytic cycle on Europa and was first detected by the Galileo Near-Infrared Mapping Spectrometer (NIMS) via a prominent absorption band at 3.5 $\micron$ \citep{Carlson1999}. The continuous bombardment of Europa's surface by electrons, protons, and ions from the Jovian magnetosphere leads to the dissociation of water molecules and the recombination of the fragments to form H$_2$O$_2$, molecular oxygen (O$_2$), and thus an oxidized surface environment \citep{JohnsonQuickenden1997,CooperEtAl2003,JohnsonEtAl2003,LoefflerEtAl2006,bain1955,Carlson1999,Carlson2009}. Surface oxidants such as H$_2$O$_2$, if transported through Europa's ice shell, could interact with ocean water that is potentially reduced by hydrothermal processes at the seafloor. Such mixing could create a redox potential, providing a viable chemical energy source for Europa's interior ocean \citep{Hand2006,Hand2009,pasek2012,Chyba2000}. Understanding what controls H$_2$O$_2$ can affect our understanding of its stability and how likely it is to get recycled into the interior. It also has implications for the production and stability of related radiolysis products, such as O$_2$ \edit1{\citep{CooperEtAl2003}}, that are produced by the same or similar processes, but harder to observe. H$_2$O$_2$, O$_2$, and H$_2$ formation from ion and electron irradiated water ice is expected to be a common radiation chemistry pathway throughout the outer solar system.

Laboratory spectral measurements and irradiation experiments have been used to understand the radiolysis and photolysis of water ice, including the production and destruction of H$_2$O$_2$ \citep[e.g.,][]{cassidy2010,HandCarlson2011,JohnsonQuickenden1997,LoefflerEtAl2006,MooreHudson2000}. In addition to water ice and sufficient irradiation being essential precursors for H$_2$O$_2$ formation, irradiation experiments have also shown an inverse relationship between the equilibrium abundances of H$_2$O$_2$ and the ice temperature \citep{HandCarlson2011,LoefflerEtAl2006,MooreHudson2000,ZhengEtAl2006}. Thus, it was expected that Europa should have an enrichment of H$_2$O$_2$ in colder, more ice-rich regions, compared to a more minimal presence in warmer, less ice-rich regions \citep{HandCarlson2011}.

\begin{table*}[ht]
\begin{sloppypar}
\begin{center}
\caption{Table of Observations\label{table:obs}}
\begin{tabular}{cccccccccc}
\hline\\[-4mm] \hline
Date & Telescope & Hemisphere & Slit & Airmass & Integration & Central & Central & Star \\ 
(UT) &  &  & Orientation & Range & (Minutes) & Longitude & Latitude & Calibrator \\ \hline
2018 Mar 29 & IRTF/Spex & Sub-Jovian$^a$ & Disk-integrated & 1.3 & 32 & 345 W & 4 S & HD 140990 \\ 
2018 Mar 29 & IRTF/Spex & Sub-Jovian$^b$ & Disk-integrated & 1.3-1.4 & 48 & 348 W & 4 S & HD 140990 \\ 
2018 May 11 & IRTF/Spex & Sub-Jovian$^c$ & Disk-integrated & 1.2-1.3 & 32 & 11 W & 4 S & HD 128596 \\ 
2018 May 11 & IRTF/Spex & Sub-Jovian$^d$ & Disk-integrated & 1.2 & 40 & 14 W & 4 S & HD 128596 \\ 
2018 Jun 30\ & IRTF/Spex & Sub-Jovian$^e$ & Disk-integrated & 1.2 & 8 & 29 W & 4 S & HD 128597 \\
2018 Jun 30 & IRTF/Spex & Sub-Jovian$^f$ & Disk-integrated & 1.2 & 43 & 31 W & 4 S & HD 128597 \\
2018 Jun 30 & IRTF/Spex & Sub-Jovian$^g$ & Disk-integrated & 1.2-1.3 & 33 & 36 W & 4 S & HD 128597 \\
2018 Jun 30 & IRTF/Spex & Sub-Jovian$^h$ & Disk-integrated & 1.3-1.5 & 38 & 39 W & 4 S & HD 128597 \\
2013 Dec 29 & Keck AO & Anti-Jovian & E/W & 1.0-1.2 & 40 & 195 W & 3 N & HD 54371 \\ 
2013 Dec 29 & Keck AO & Anti-Jovian & E/W & 1.1 & 20 & 195 W & 25 S & HD 54372 \\ 
2016 Feb 24 & Keck AO & Trailing/sub-Jovian & E/W & 1.5-1.7 & 40 & 334 W & 2 S & HD 98947 \\ 
2016 Feb 24 & Keck AO & Trailing/sub-Jovian & E/W & 1.2-1.3 & 40 & 338 W & 0 N & HD 98947 \\ 
2016 Feb 24 & Keck AO & Trailing/sub-Jovian & E/W & 1.0-1.1 & 40 & 342 W & 1 S & HD 98947 \\ 
2016 Feb 24 & Keck AO & Sub-Jovian & N/S & 1 & 40 & 350 W & 2 S & HD 98947 \\ 
2016 Feb 24 & Keck AO & Sub-Jovian & N/S & 1 & 40 & 348 W & 2 S & HD 98947 \\ 
2016 Feb 25 & Keck AO & Leading & E/W & 1.2-1.4 & 40 & 79 W & 0 S & HD 98947 \\ 
2016 Feb 25 & Keck AO & Leading & E/W & 1.1-1.2 & 40 & 82 W & 12 S & HD 98947 \\ 
2016 Feb 25 & Keck AO & Leading & N/S & 1.0-1.1 & 40 & 88 W & 2 S & HD 98947 \\ 
2016 Feb 25 & Keck AO & Leading & E/W & 1.1 & 40 & 95 W & 0 N & HD 98947 \\ 
2018 Jun 06 & Keck AO & Leading & E/W & 1.5 & 20 & 115 W & 7 S & HD 128596 \\ 
2018 Jun 06 & Keck AO & Leading & N/S & 1.4 & 10 & 120 W & 5 S & HD 128596 \\ 
2018 Jun 06 & Keck AO & Leading & N/S & 1.4 & 10 & 126 W & 5 S & HD 128596 \\ 
2018 Jun 06 & Keck AO & Leading/anti-Jovian & N/S & 1.2-1.3 & 20 & 171 W & 2 S & HD 128596 \\ 
2018 Jun 06 & Keck AO & Leading/anti-Jovian & N/S & 1.2-1.3 & 20 & 150 W & 2 S & HD 128596 \\ 
2018 Jun 06 & Keck AO & Leading/anti-Jovian & N/S & 1.3 & 10 & 166 W & 1 S & HD 128596 \\ 
2018 Jun 06 & Keck AO & Leading & N/S & 1.2 & 15 & 113 W & 4 S & HD 128596 \\ 
2018 Jun 06 & Keck AO & Leading & N/S & 1.2-1.3 & 20 & 100 W & 6 S & HD 128596 \\ 
2018 Jun 06 & Keck AO & Leading & N/S & 1.3 & 20 & 82 W & 3 S & HD 128596 \\ 
2018 Jun 06 & Keck AO & Leading & E/W & 1.3-1.4 & 20 & 131 W & 1 S & HD 128596 \\ 
2018 Jun 07 & Keck AO & Trailing & E/W & 1.2-1.3 & 40 & 222 W & 4 S & HD 128596 \\ 
2021 Sep 24 & Keck AO & Trailing & E/W & 1.2-1.4 & 60 & 275 W & 20 N & HD 203311 \\ 
2021 Sep 25 & Keck AO & Trailing & N/S & 1.3-1.4 & 40 & 318 W & 8 N & HD 203311 \\ 
2021 Sep 25 & Keck AO & Sub-Jovian/Leading & N/S & 1.3 & 40 & 41 W & 2 N & HD 203311 \\ 
2022 Nov 23 & JWST & Leading & - & - & 21 & 93 W & 3 N & GSPC P330-E \\ \hline
\end{tabular}
\end{center}
\end{sloppypar}

[a] 2 hours pre-eclipse; [b] 1 hour pre-eclipse; [c] 1 hour post-eclipse; [d] 2 hours post-eclipse; [e] 3 hours post-eclipse; [f] 4 hours post-eclipse; [g] 5 hours post-eclipse; [h] 6 hours post-eclipse

\end{table*}

However, contrary to the laboratory predictions, recent Keck AO observations of Europa's 3.5-$\micron$ H$_2$O$_2$ absorption feature across its leading hemisphere showed enhanced H$_2$O$_2$ in geologically young, low-latitude chaos terrains characterized by lower water-ice abundance \citep{BrownHand2013, Fischer2015, Fischer2016}, warmer temperatures \citep{Rathbun2010,Trumbo_2018}, and a notable enrichment of sodium chloride (NaCl) \citep{Trumbo2019nacl,Trumbo2022}. This observation finds contradictory results on the effects of both water-ice availability and temperature. While the findings indicate that temperature is not the dominant factor, it remains uncertain whether temperature plays a role at all in governing Europa's H$_2$O$_2$. 

In fact, the distribution suggests that the underlying composition of the chaos terrain exerts a stronger control on H$_2$O$_2$ abundance. Thus, the authors hypothesized that carbon dioxide (CO$_2$) may be enhanced in these same terrains \citep{Trumbo2019h2o2}, where it may act to enhance H$_2$O$_2$ concentrations \edit1{by inhibiting the breakdown of newly formed H$_2$O$_2$ by consuming the destructive electrons produced as the ice continues to be irradiated \citep{MooreHudson2000}}---an effect supported at the time by only limited laboratory data \citep{MooreHudson2000,Strazzulla2005} and unpublished Galileo NIMS spectra \citep{Carlson2001}. However, this hypothesis remains untested due to the lack of reliable CO$_2$ mapping on Europa and the inability to observe it from the ground, leaving the effects of temperature and composition on Europa's H$_2$O$_2$ largely unresolved.

To further investigate the effects of temperature and CO$_2$ on Europa's H$_2$O$_2$ concentrations, we analyze a combination of ground- and space-based observations of Europa's H$_2$O$_2$ from the the NASA Infrared Telescope Facility (IRTF), Keck Observatory, and JWST. Using IRTF/SpeX, we \edit1{take disk-integrated observations} Europa’s 3.5-$\mu$m H$_2$O$_2$ absorption before and after eclipse, which preserves approximately the same geometry across a large temperature change, enabling a largely controlled test of possible temperature effects. To examine the reaction of H$_2$O$_2$ to temperature fluctuations across longer, diurnal timescales, we present spatiotemporally resolved observations across Europa's day acquired using the Keck II telescope's Near InfraRed Spectrograph (NIRSPEC) and adaptive optics (AO) system. Finally, we use \edit1{spatially-resolved} JWST Cycle 1 NIRSpec observations of Europa, which simultaneously span wavelengths sensitive to both H$_2$O$_2$ and CO$_2$, to search for the previously hypothesized correlation between these two species.

\section{Observations and Data Reduction} \label{sec:observations}
\subsection{Disk-integrated Pre- and post-eclipse IRTF/SpeX observations} \label{sec:observations}

To disentangle the effect of temperature on Europa’s H$_2$O$_2$ from that of composition, we designed a controlled temperature experiment using disk-integrated spectra obtained in one-hour windows across two hours pre-eclipse and six hours post-eclipse with SpeX \citep{Rayner2003} at the NASA Infrared Telescope Facility (IRTF) across several dates \edit1{with similar eclipse windows} in 2018. \edit1{The SpeX data obtained span a wavelength range of 1.68–4.23 $\mu$m with R $\sim$ 2,000.} Dates, time, and geometries are given in Table \ref{table:obs}. 

Depending on the exact thermal inertia of the surface materials, an eclipse leads to a sudden temperature drop of 10s of degrees Kelvin, while the satellite maintains approximately the same geometry as viewed from Earth. (Figure \ref{fig:irtf}A). As simple extrapolation from laboratory studies suggests a $\sim$20 K temperature drop may increase the equilibrium band area of the H$_2$O$_2$ absorption by approximately a factor of $\sim$4 \citep{HandCarlson2011}, we assessed changes in Europa’s H$_2$O$_2$ absorption band immediately before and after eclipse and throughout this recovery period to look for such a response. As Europa is tidally locked to Jupiter and experiences an eclipse of its sub-Jovian hemisphere every orbit, pre- and post-eclipse spectra taken on different dates are still sampling the same temperature drop and recovery (Table \ref{table:obs}). Our data cover the wavelength range of $\sim$1.68-4.2 $\mu$m. For pre-eclipse observations, we observed HD 140990, a V = 7.85 G2V star with a 7.08\degree{} separation from Europa, as the telluric calibrator. For post-eclipse observations, we observed HD 128596, a V = 7.48 G2V star with a 2.45\degree{} separation from Europa, as the telluric calibrator. 

We used Spextool (Spectral Extraction TOOL), an Interactive Data Language (IDL)-based data reduction package \citep{Cushing2004}, to reduce the IRTF/SpeX data. We followed the standard methodology of flat field correction, image pair subtraction, target detection, spectra extraction, wavelength calibration, combination of multi-order spectra, and telluric correction \citep{Vacca_2003}. Spectra from within each 1-hour window were merged into one single spectrum per window. We clipped out bad pixels and any highly variable telluric lines by comparing data points to the means of a 51-point window with a threshold of 0.01 \edit1{, as determined by manually inspecting the spectra so that only telluric lines and outliers are clipped}. \edit1{To further improve the signal-to-noise ratio, we then smoothed the spectra using a Savitzky-Golay filter with a window length of 5 data points and a polynomial order of 3}.

Though the geometry was nearly identical between subsequent time windows, Europa did rotate $\sim$50 degrees in longitude across the entire 8 hours of observation. To address this caveat, we used spatially resolved Keck AO observations of the sub-Jovian hemisphere to estimate the effects of small changes in observed band strength attributable to the minor amount of rotation in the eclipse experiment.

\subsection{Spatially resolved Keck AO observations across Europa's day} \label{sec:observations}
The creation and destruction of H$_2$O$_2$ on Europa exists in a dynamic equilibrium, and the timescale for response to a temperature perturbation is unkown. Indeed, the factor of $\sim$4 change in H$_2$O$_2$ band strength
from laboratory studies mentioned above is a direct comparison of H$_2$O$_2$ band strengths in pure water ice at different temperatures irradiated until equilibrium concentrations were achieved \citep{HandCarlson2011}, which is not directly translatable to Europa. Thus, we also evaluate the response of H$_2$O$_2$ to temperature variations across a longer timescale than eclipse by using Keck NIRSPEC and the AO system on the Keck II telescope to observe Europa's 3.5-$\micron$ H$_2$O$_2$ band throughout its day. 

We observed the same locations on rotational timescales from 4 hours (comparable in timeframe to Europa's eclipse recovery) to 29 hours ($\sim$1/3 Europa day and equivalent to progression from morning to afternoon) and assessed their localized diurnal variability. In addition, we observed some locations that expand on past maps of the spatial distribution of Europa's H$_2$O$_2$ \citep{Trumbo2019h2o2}. Observations used in this analysis were acquired in 2013, 2016, 2018, and 2021. Dates, times, and geometries are given in Table \ref{table:obs}. We used the 3.96$^\prime$$^\prime$ × 0.072$^\prime$$^\prime$ slit in low-resolution mode (R = 2000) across the L band wavelengths of approximately 3.1–4 $\mu$m. 

In 2013, we used HD 54371, a V = 7.1 G6V star at 3.2\degree{} separation from Europa, as the telluric calibrator. Each telluric calibrator pointing consisted of 1 12-20s coadd or 2 15s coadds. In 2016, we used HD 98947, a V = 6.9 G5 star at 1.2\degree{} separation from Europa, as the telluric calibrator. Each telluric calibrator pointing consisted of 2 15s coadds. In 2018, we used HD 128596, a V = 7.5 G2V star with a 3.9\degree{} separation from Europa. Each telluric calibrator pointing consisted of 2 10s coadds. In 2021, we used HD 203311, a V = 7.45 G2V star with a 5.15\degree{} separation from Europa. Each telluric calibrator pointing consisted of 2 15s coadds.

We obtained ephemeris data and viewing geometry from JPL Horizons. During the observations, Europa's apparent diameter was approximately 1$^\prime$$^\prime$, translating to about 10 resolution elements ($\sim$300 km resolution at the sub-observer point) at the diffraction limit of Keck at 3.5 $\mu$m. We shifted the pointing of targets between opposite ends along the slits in an AB, ABBA, or ABBAAB pattern, to enable pair-subtraction during reduction. Corresponding slit-viewing camera (SCAM) images were acquired to ensure a consistent slit position across each observation set and to allow for later determination of the corresponding geographic coordinates on Europa.

Some of the observations displayed readout artifacts (artificial brightening of every 8 or 64 pixels in the dispersion direction), which we replaced with the mean of the two immediately adjacent pixels. Following the data reduction outlined in \citet{Trumbo2019h2o2}, we then used Python packages (astropy \citep{astropy2013,astropy2018,astropy2022}, scikit-image \citep{scikitimage2014}, scipy \citep{SciPy2020}, cartopy \citep{Cartopy2015}, astroquery \citep{astroquery2019}) in following the standard reduction steps of image rectification, flat field correction, image pair subtraction, residual sky subtraction, spectra extraction, and telluric calibration. We used an ATRAN atmospheric transmission spectrum \citep{lord1992} for wavelength calibration. The first several columns of the flatfields (separate from the target's location on the chip) were consistently unevenly illuminated, so we replaced the unilluminated edge with the mean of the adjacent illuminated window of the same width in pixels.

We extracted summed spectra in 8-pixel resolution elements, stepping by 4 spatial pixels. As with the IRTF spectra, we clipped out bad pixels and any highly variable telluric lines by comparing data points to the means of a 51-point window with a threshold adjusted for each observation. Spectra were then smoothed using a Savitzky-Golay filter with an optimal window length and a polynomial order of 3 to improve signal-to-noise.

To estimate the coordinates of the slit on Europa and extract spectra for each resolution element, we aligned SCAM images to each NIRSPEC exposure. With detector resolutions of 0.013$^\prime$$^\prime$/pixel (pre-upgrade NIRSPEC), 0.009$^\prime$$^\prime$/pixel (updated NIRSPEC after 2018), 0.0168$^\prime$$^\prime$/pixel (pre-upgrade SCAM), and 0.0149$^\prime$$^\prime$/pixel (updated SCAM), we calculated the size of Europa in both NIRSPEC and SCAM pixels. We then located Europa in the 2D spectral images and estimated the coordinates of the slit by calculating the slit pixel offsets from Europa's center in the SCAM images. With the known NIRSPEC pixel scale, we determined the geographic coordinates of each resolution element. \edit1{When combining individual exposures (from A and B nods), we account for small changes to the effective slit width that result from minor shifts in the pointing on Europa. We estimate the spatial uncertainty from our SCAM navigation procedure to be one SCAM pixel, which is much smaller than the $\sim$0.09$^\prime$$^\prime$ diffraction limit of Keck at 3.5 $\mu$m. We use the extracted coordinates to map} the geographic distribution of the 3.5-$\mu$m H$_2$O$_2$ band. For overlapping regions, we averaged band strengths. We excluded data from limb pixels in the trailing hemisphere slits, where the data quality was too poor to reliably measure the H$_2$O$_2$ band.

\subsection{Spatially resolved JWST observation} \label{sec:observations}
As CO$_2$ has been proposed as a potential factor enhancing H$_2$O$_2$ concentrations in chaos terrains \citep{Trumbo2019h2o2,Carlson2001,Carlson2009}, we assessed the correlation between the geographic distribution of Europa's H$_2$O$_2$ and CO$_2$ using archival JWST NIRSpec data (from the Cycle 1 Guaranteed Time Observations program \#1250) of Europa's leading hemisphere, observed on November 23, 2022, at 08:18 UT. The data we analyzed were obtained with the NIRSpec integral field unit (IFU), the high-resolution gratings (G235H and G395H, R = 2700), and the F170LP and F290LP filters at wavelengths of approximately 1.7–5.2 $\mu$m. Only Europa's leading hemisphere (sub-observer longitude of 93\degree{}W) was observed during this program. We used a spectrum of GSPC P330-E (Program 1538 \citep{Gordon_2022}), a G0V star, as a solar analog. We followed the same data reduction steps as in \citet{TrumboBrown2023} and \citet{trumbo2023_h2o2Ganymede} for both the Europa data and the stellar data, and extracted spectra corresponding to each 0.1$^\prime$$^\prime$ x 0.1$^\prime$$^\prime$ spatial pixel on Europa. 

\begin{figure*}[ht]
\figurenum{1}
\plotone{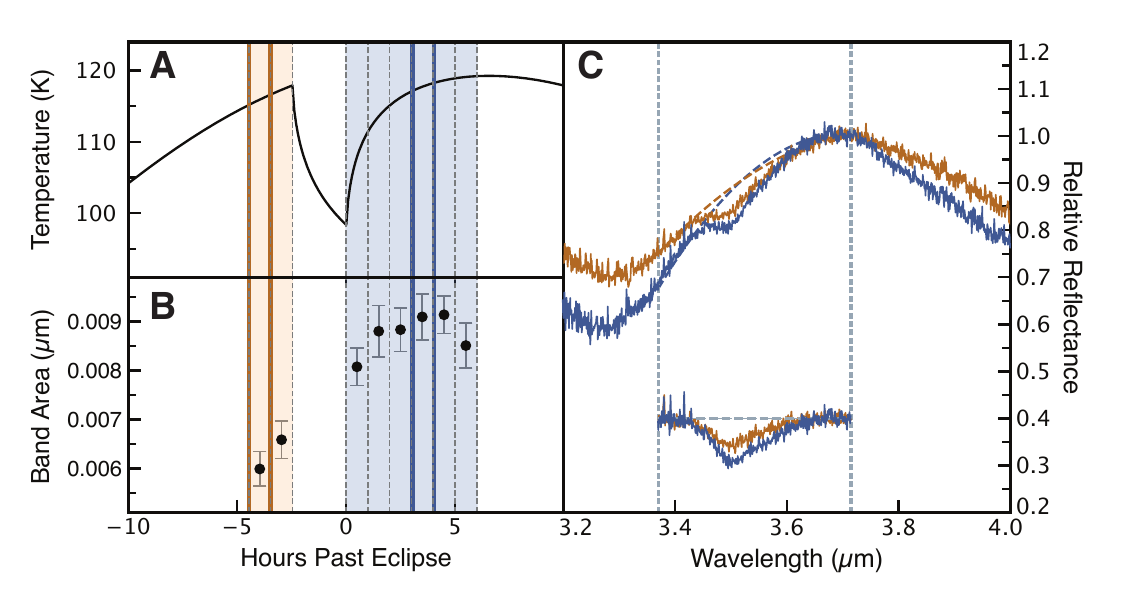}
\caption{[A] The temperature of sub-Jovian Europa as it passes through eclipse. This temperature profile is based on the thermal model of \citep{Trumbo2017alma}.  Sub-Jovian Europa was observed for 2 hours before eclipse and 6 hours after eclipse (no observations during eclipse due to the lack of reflected light). The bottom of the drop indicates the end of eclipse. The sub-Jovian temperature takes $\sim$6 hours to recover from the estimated $\sim$20 K drop during eclipse. Spectra were averaged for each of the $\sim$1-hour time intervals. Orange indicates pre-eclipse observations and blue indicates post-eclipse observations. [B] Integrated band area calculated for each $\sim$1-hour window. A factor of $<$2 change is observed both between the measurements immediately pre-eclipse and immediately post-eclipse, and between the weakest 3.5-$\mu$m absorption at 2 hours pre-eclipse and the strongest 3.5-$\mu$m absorption 5 hours post-eclipse. This small change is attributable to rotation of Europa across this time period and rules out a strong response to the eclipse temperature drop (discussed in section \ref{sec:results1}). As errors associated with the unknown background continuum dominate over those due to noise, we estimate error bars as the values given using a range of plausible continuum fits across the wavelength range from 3.33–3.37 $\mu$m to 3.715–3.78 $\mu$m. [C] Example \edit1{IRTF/Spex} spectra comparing the 3.5-$\mu$m absorption 2-hours pre-eclipse with the absorption 4-hours post-eclipse. The boundaries of the selected time windows are bold in [B]. Dashed gray lines outline the H$_2$O$_2$ band. Second-order polynomial continua are indicated by the dashed curves of the same color as the spectra. Continuum-removed absorptions are included to ease comparison of the band strengths. \edit1{The depth of the 3-$\mu$m water ice absorption band (shown here as the 3.2–3.7 $\mu$m range following the Fresnel peak) appears to increase after eclipse (blue)}, suggesting a possible rotational effect as Europa rotates from a more trailing/sub-Jovian geometry towards a more sub-Jovian/leading geometry. \label{fig:irtf}}
\end{figure*}

\subsection{Band strength calculation} \label{sec:observations}
To assess the response of Europa's H$_2$O$_2$ to temperature changes and its potential correlation with CO$_2$, we measure the strength of the 3.5-$\mu$m H$_2$O$_2$ feature in all extracted spectra from IRTF, Keck, and JWST. For all spectra, we excluded the absorption feature itself (3.4–3.65 $\mu$m) before fitting a second-order polynomial across the wavelength range of 3.37 to 3.715 $\mu$m. Each fit was visually inspected, and minor adjustments were made as needed to ensure an optimal fit to the continuum. We then divided the fitted continuum from each spectrum to isolate the absorption feature. Band areas (i.e. equivalent widths) were calculated by integrating the area of the residual absorption.

CO$_2$ displays three signatures in the JWST data---a narrow band near 2.7 $\micron$ and two minima within the $\nu$$_3$ band at $\sim$ 4.25 $\micron$ and 4.27 $\micron$ \citep{TrumboBrown2023, Villanueva2023}. To assess the correlation with our H$_2$O$_2$ measurements, we used published band areas of the $\nu$$_3$ band from \citep{TrumboBrown2023}, separately calculated the band depths of each $\nu$$_3$ minimum relative to the continua fit by those authors, and calculated the area of the 2.7-$\micron$ band by simply fitting and removing a linear continuum. We excluded some pixels close to Europa’s limb that appeared particularly affected by flux-oscillation artifacts resulting from the instrument's undersampling of the point spread function. These oscillations significantly impact areas where Europa’s signal changes rapidly \citep{TrumboBrown2023}. Oscillations affected the measurement of the 3.5-$\mu$m H$_2$O$_2$ feature in more pixels than that of the CO$_2$ feature, which resulted in the exclusion of a few more pixels than \citet{TrumboBrown2023}.

\begin{figure*}[ht]
\figurenum{2}
\plotone{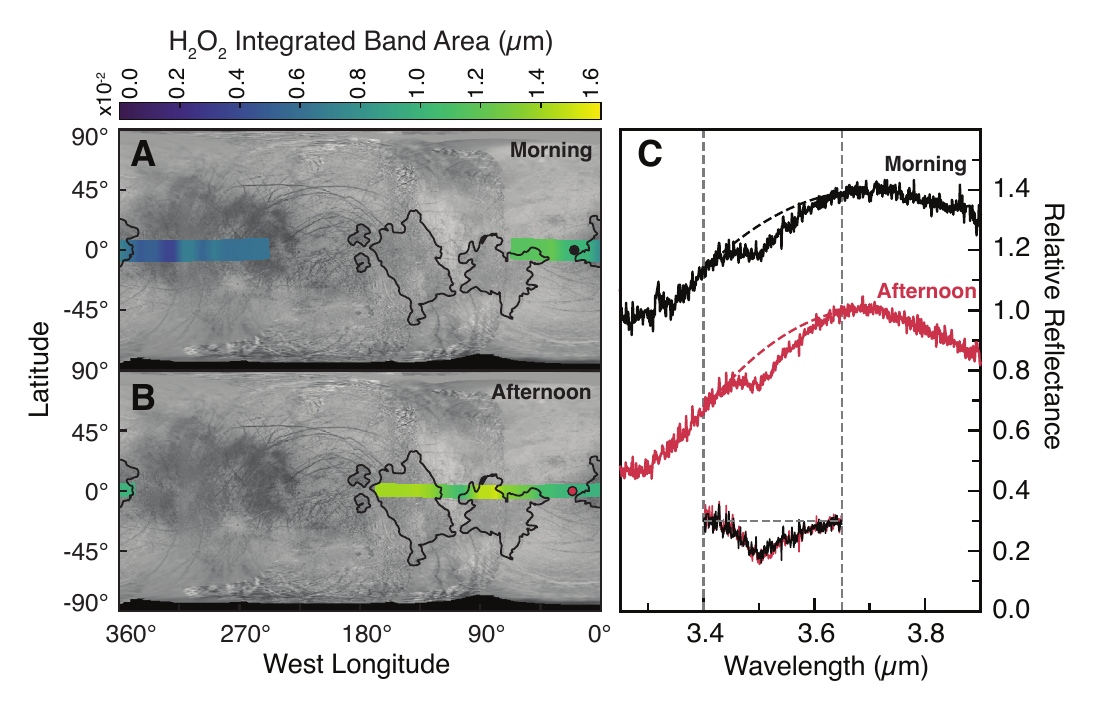}
\caption{[A] Example \edit1{Keck NIRSPEC} slit obtained on February 24, 2016, when 18.6\degree{}W (labeled with the black dot) is in the morning (\edit1{$\sim$13 hours prior to local noon in Europa's $\sim$85-hour day,} estimated $\sim$103 K). [B] Example \edit1{Keck NIRSPEC} slit obtained on February 25, 2016, when the same region (18.5\degree{}W; labeled with the red dot) is in the afternoon (\edit1{$\sim$16 hours after local noon in Europa's $\sim$85-hour day,} estimated $\sim$115 K). [C] Corresponding reflectance spectra for this location in the morning (in black) and in the afternoon (in red). Dashed gray lines outline the H$_2$O$_2$ band. Second-order polynomial continua are indicated by the dashed curves of the same color as the spectrum. Continuum-removed absorptions are included and show nearly overlapping absorptions. A $\sim$12 K temperature increase over 24 hours is estimated from the thermal model of \citet{Trumbo2017alma}. This temperature change results in a negligible -3.1$\%$ change in the integrated band areas. Leading chaos regions are outlined in black \citep{leonard2024}. Background basemap credit: \citet{USGSbasemap}.\label{fig:keck}}
\end{figure*}

\section{Pre- and post-eclipse H$_2$O$_2$ Measurements} \label{sec:results1}
To isolate the effect of temperature on Europa’s H$_2$O$_2$ from those of composition or geography, we measure the 3.5-$\mu$m H$_2$O$_2$ band on Europa's sub-Jovian hemisphere across a 2-hour period before eclipse and across a 6-hour period following the conclusion of its daily eclipse. Using the thermal model of \citet{Trumbo2017} and the best-fit thermal inertia value (60 J/(m\textsuperscript{2}Ks\textsuperscript{1/2})) found in \citet{Trumbo_2018} from ALMA observations of the sub-Jovian hemisphere, we estimate an expected $\sim$20 K temperature drop across eclipse and a recovery to pre-eclipse temperatures that span the 6-hour post-eclipse period across which we observed. (Figure \ref{fig:irtf}A). 

\citet{HandCarlson2011} irradiated pure water ice with 10 keV electrons at temperatures of 80 K, 100 K, and 120 K and monitored the change in H$_2$O$_2$ band absorption until equilibrium concentrations were achieved at flux levels of $\sim$$10^{19}$ eV cm$^{-2}$. The authors' measured equilibrium H$_2$O$_2$ band area at 120 K is approximately a factor of 4 weaker than the measured equilibrium H$_2$O$_2$ band area at 100 K and a factor of 6 weaker than the measured equilibrium H$_2$O$_2$ band area at 80 K. This laboratory result is consistent with the expected increased mobility of OH with increasing temperature \citep{JohnsonQuickenden1997,HandCarlson2011,LoefflerEtAl2006,quikenden1991}. \edit1{Though, it is worth noting that the magnitude of temperature effects seen in laboratory studies varies, with some experiments \citep[e.g.,][]{GomisLeto2004,LoefflerEtAl2006, ZhengEtAl2006} observing smaller changes in H$_2$O$_2$ concentration than \citet{HandCarlson2011} and some modeling work predicting smaller effects as well \citep{Teolis2017}. Nevertheless,} taking the most simplistic extrapolation from these experiments, we might expect up to a factor of 4 increase in our observed H$_2$O$_2$ band area immediately after the $\sim$20-K eclipse temperature drop and then a slow decrease of the absorption band as the temperature climbs back up toward pre-eclipse values.

In Figure \ref{fig:irtf}B, we plot H$_2$O$_2$ band areas of the averaged spectra from each 1-hour time window surrounding the eclipse, where the error bars are estimated by varying the continuum fits across the wavelength range from 3.33–3.37 $\mu$m to 3.715–3.78 $\mu$m. The unknown true continuum presents a larger source of error than noise, but our error bars are still far smaller than the large increases in band strength we seek to detect. All of the spectra used to derive the band areas in Figure \ref{fig:irtf} are included in Figure \ref{fig:supp_irtf}. We find that our data clearly rule out such a dramatic increase in the H$_2$O$_2$ band following eclipse, instead exhibiting only a small 22.6\% change (factor of 1.23) between the measurements immediately pre-eclipse and immediately post-eclipse, and a 38.7\% change (factor of 1.53) between the weakest 3.5-$\mu$m absorption at 2 hours pre-eclipse and the strongest 3.5-$\mu$m absorption 5 hours post-eclipse.

It is possible that this small change does reflect a delayed response to the eclipse temperature drop resulting from the unknown timescale needed for Europa's surface H$_2$O$_2$ concentrations to re-equilibrate. However, the 2- and 3-$\micron$ water-ice absorption bands also appear to change following the eclipse and across the recovery period, which suggests that Europa's small degree of rotation from the slightly less-icy trailing/sub-Jovian geometry pre-eclipse (central longitude = 348\degree{}W) towards a slightly more-icy sub-Jovian/leading geometry (central longitude = 36\degree{}W) by 5 hour post-eclipse needs to be considered as well. Our spatially resolved Keck observation across the sub-Jovian hemisphere suggests up to a factor of $\sim$2.1 increase in H$_2$O$_2$ band strength as the longitude shifts from 352\degree{}W to 33\degree{}W degrees. Thus, the small change (factor $<$2) of H$_2$O$_2$ band strength following eclipse could be reasonably attributed to the $\sim$50 degrees of rotation of Europa surrounding eclipse, and our eclipse experiment reveals no clear evidence for any response of Europa's H$_2$O$_2$ to the temperature changes across eclipse, let alone the large factors seen in laboratory experiments. However, it is possible that the unknown time it would take Europa's H$_2$O$_2$ to re-equilibrate following such a perturbation is longer than the eclipse experiment allowed.

\section{Keck H$_2$O$_2$ Diurnal Variability} \label{sec:results2}

While we find no clear effect of temperature in our IRTF eclipse experiment, the laboratory data that see strong differences in H$_2$O$_2$ with temperature represent a direct comparison of equilibrium concentrations \citep[e.g.,][]{HandCarlson2011}, which may not have been reached across the eclipse timescale. For this reason, we also use Keck NIRSPEC AO observations to look for variability in Europa's H$_2$O$_2$ over longer, diurnal timescales. In particular, we examine slits oriented in an east/west direction that cover the same surface regions at different times of the local day. As noted above, \citet{HandCarlson2011} suggest that equilibrium abundances of H$_2$O$_2$ can vary by a factor of 4 across the 100-120 K temperature range relevant to Europa and that H$_2$O$_2$ can be destroyed during the day in equatorial regions and produced efficiently at night across the surface, such that higher H$_2$O$_2$ concentration might be expected in the early morning as opposed to in the warm afternoon.

During our 2016 and 2018 observations, we purposely aligned slits taken on adjacent nights to investigate the hypothesized diurnal variability of H$_2$O$_2$. We examine 5 east-west slit pairs centered within $\pm$20$\degree$ of the equator, spanning a timescale of 20-29 hours—approximately one-quarter to one-third of a Europa day—and covering the temperature progression from morning to afternoon. Similar to what we see from the eclipse experiment, the changes in integrated band area from morning to afternoon appear to be much smaller than the equilibrium variations observed in the lab and instead range from -32$\%$ to 30$\%$ (factor of 0.68–1.3), with no clear or consistent trend attributable to diurnal variability. 

As a representative example, in Figure \ref{fig:keck}, we compare spectra of the same location (18.6\degree{}W, 0.4\degree{}N) observed in the morning (cold temperatures) and afternoon (warmer temperatures) of February 2016. Using the thermal model of \citet{Trumbo2017alma} and a thermal inertia of 60 J/(m\textsuperscript{2}Ks\textsuperscript{1/2}) suggested by ALMA observations of the sub-Jovian hemisphere \citep{Trumbo_2018} and consistent with thermal inertia values estimated from the Galileo Photopolarimeter-Radiometer (PPR) observations \citep{Spencer1999, Rathbun2010}, we estimate a corresponding $\sim$12 K temperature increase over a $\sim$24-hour timescale for this location. We observe only a minimal change (-3.1$\%$) in the integrated H$_2$O$_2$ band areas for this location, and, indeed, the entirety of the overlapping portions of the two slits appear consistent. All of the overlapping E/W slit positions we inspect are presented in Figure \ref{fig:supp}.

To expand on this analysis, we also assess 2 east/west slit pairs observed over a timescale of 4 and 16 hours between 2013 and 2018 (rather than subsequent nights) in our larger dataset (Figure \ref{fig:supp}). We observe only minimal changes in integrated band areas from -0.24$\%$ to 0.44$\%$ (factor of 0.76-1.44). We also note that there appears to be some evidence for stochastic H$_2$O$_2$ variability between years, as noted in \citet{Trumbo2019h2o2}, where the bands observed in 2016 are overall stronger than in later years (Figure \ref{fig:supp}). We suggest that this could be related to potential variability of the dynamic radiation environment.

The lack of a strong or consistent diurnal trend, along with the minimal changes observed across eclipse, suggest that Europa's H$_2$O$_2$ is largely insensitive to the temperature fluctuations that it experiences on a daily basis. Thus, our results indicate that, in addition to temperature not influencing the spatial variations of Europa's H$_2$O$_2$, temperature also appears not to drive strong temporal variations. We discuss this apparent discrepancy with laboratory work further in section \ref{sec:discussion}.

\section{Keck H$_2$O$_2$ Mapping} \label{sec:results2}
\begin{figure}[ht]
\figurenum{3}
\plotone{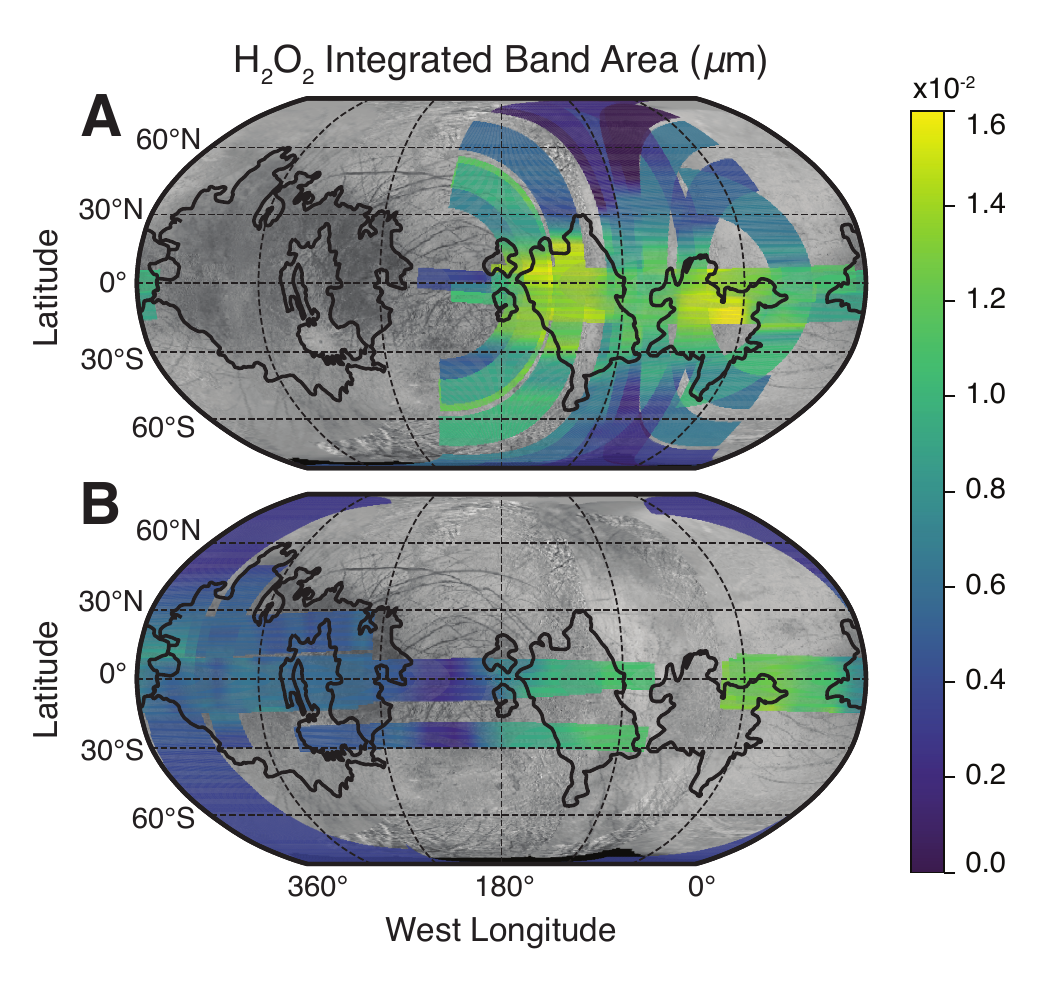}
\caption{Mapped \edit1{Keck NIRSPEC} slits from our 2013, 2016, 2018, and 2021 observations on the leading [A] and trailing [B] hemispheres, which show the spatial distribution of Europa's 3.5-$\mu$m H$_2$O$_2$ band. Consistent with full-disk spectra \citep{HandBrown2013} and previous spatially resolved spectra \citep{Trumbo2019h2o2}, the geologically young chaos regions on the leading hemisphere \citep[outlined in black;][]{leonard2024} exhibit the largest H$_2$O$_2$ features. In contrast, the trailing hemisphere demonstrates generally much weaker absorptions compared to the leading hemisphere and a depletion of H$_2$O$_2$ west of the anti-Jovian point around 210°W,  corresponding to a gap between the large-scale chaos regions of the leading and trailing hemispheres. We note that observations from various years are mapped together to illustrate the updated coverage of Europa's H$_2$O$_2$, despite some apparently stochastic fluctuations in H$_2$O$_2$ between 2016 and later years. Background basemap credit: \citet{USGSbasemap}.\label{fig:keck_map}}
\end{figure}

In addition to suggesting that diurnal temperature changes do not strongly control Europa's H$_2$O$_2$, the Keck observations provide additional spatial coverage of Europa's surface compared to past maps of H$_2$O$_2$ \citep{Trumbo2019h2o2}. In Figure \ref{fig:keck_map}, we present maps of H$_2$O$_2$ band area composed of east/west orientated and north/south oriented slits from 2013, 2016, 2018, and 2021, which include additional coverage beyond the previous maps. We separated out coverage of the leading and trailing hemispheres for clarity\edit1{, and to highlight the longitudinal variation along the two included anti-Jovian E/W slits}. Overall, the updated distribution agrees well with the trends shown by \citet{Trumbo2019h2o2}.

Our maps present a clear difference between low-latitude chaos and low-latitude plains terrain in east/west slits crossing the leading chaos regions (Figure \ref{fig:keck}B, Figure \ref{fig:keck_map}A, Figure \ref{fig:supp}A-D), and a north/south slit just east of Tara Regio ($\sim$85° W) (Figure \ref{fig:keck_map}A) shows overall weaker H$_2$O$_2$ absorption and less latitudinal variation in  H$_2$O$_2$ than do north/south slits crossing Tara Regio. This distribution further confirms the curious enrichment of the 3.5-$\mu$m H$_2$O$_2$ feature within the large-scale chaos terrains at low-to-mid-latitudes on the leading hemisphere \citep{Trumbo2019h2o2}, which are understood to contain endogenic non-ice material like salts \citep{Fischer2015, Fischer2016, Trumbo2019nacl, Trumbo2022}.

Our maps also provide additional north/south and east/west coverage on the trailing hemisphere. Given the drastic changes in the spectral continuum shape between the two hemispheres associated with changes in water-ice abundance, we carefully inspected all continuum-fitting manually and adjusted fitting parameters accordingly. Consistent with the hemispheric data of \citet{HandBrown2013} and the spatially resolved data of \citet{Trumbo2019h2o2}, the trailing regions show much weaker H$_2$O$_2$ absorption overall. This hemispheric difference in H$_2$O$_2$ abundance may stem from the strong contrast in water ice fraction between the leading and trailing hemispheres \citep{Trumbo2019h2o2,HandBrown2013}. The widespread SO$_2$ on the trailing hemisphere \citep[e.g.,][]{Becker2022} may also contribute to the depletion of H$_2$O$_2$ via interactions to create sulfate \citep{Trumbo2019h2o2,LoefflerHudson2013}.

As the H$_2$O$_2$ absorption on the trailing hemisphere is so weak against the background continuum, we cannot robustly distinguish between trailing chaos and non-chaos regions or make a clear detection of any latitudinal trend on this hemisphere. However, we do repeatedly see a depletion of H$_2$O$_2$ west of the anti-Jovian point around 210°W, which corresponds to a gap between the large-scale chaos regions of the leading and trailing hemispheres (Figure \ref{fig:keck_map}, Figure \ref{fig:supp}), hinting that similar geological associations could exist below our detection across the trailing side. This depletion around 210$\degree$ W also corresponds to lower mean reflectances across the L-band (3.15 to 3.95 $\mu$m) mapped in \citet{Fischer2016}. Future, high-signal-to-noise JWST observations of Europa’s trailing hemisphere would help further investigate the geographic variation of its weak H$_2$O$_2$ bands.

With the additional Keck coverage, we confirm a strong correlation between concentrated H$_2$O$_2$ and chaos terrains. This suggests that the endogenic composition of these regions may play an important role in controlling the distribution of H$_2$O$_2$ and that a possible correlation with CO$_2$ remains a key hypothesis \citep{Trumbo2019h2o2}.

\section{JWST Mapping and Correlation with CO$_2$} \label{sec:results3}
To investigate the hypothesized effect of CO$_2$ on the unexpected distribution of H$_2$O$_2$ \citep{Trumbo2019h2o2}, we measure the H$_2$O$_2$ band in the Cycle 1 JWST data recently used to map Europa's leading-hemisphere CO$_2$ \citep{TrumboBrown2023,Villanueva2023} and evaluate whether H$_2$O$_2$ and CO$_2$ are geographically correlated. Both features are observed simultaneously in the same dataset, ensuring the best geographic alignment.

\begin{figure}[ht]
\figurenum{4}
\plotone{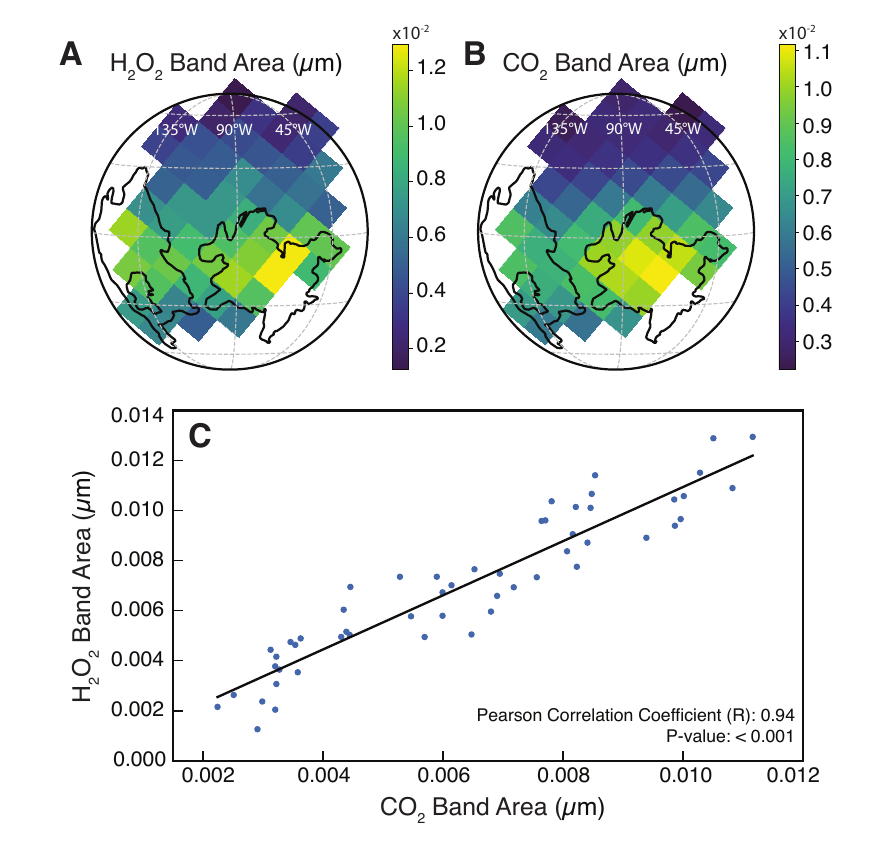}
\caption{[A] Map of the H$_2$O$_2$ integrated band area from the JWST Cycle 1 observations. Gray dashed lines indicate the meridians and the 60° S, 30° S, 0° N, 30° N, and 60° N parallels. The strongest absorption occurs in Tara Regio ($\sim$85° W) to the right of the 90° W meridian and in parts of the chaos region Powys Regio ($\sim$125° W) on the left portion of the disk. [B] Map of the entire $\nu$$_3$ CO$_2$ integrated band area ($\mu$m) \citep{TrumboBrown2023}. Concentrated CO$_2$ occurs in the same regions as concentrated H$_2$O$_2$---the leading chaos terrains. The disk of Europa is indicated by a black circle centered at 2.7°N, 93°W. [C] Linear fit illustrating the relationship between the distributions of the two features. A Pearson correlation coefficient of 0.94 (P\(<\)0.001) signifies a strong positive correlation between the distributions of H$_2$O$_2$ and CO$_2$ on the leading hemisphere. Consistent with \citet{TrumboBrown2023}, limb pixels affected by the instrument's inherent undersampling of the point spread function are excluded. Oscillations affected the measurement of the 3.5-$\mu$m H$_2$O$_2$ feature in more pixels than that of the CO$_2$ feature. Therefore, a few more pixels are excluded than in \citet{TrumboBrown2023}. Leading chaos regions are outlined in black \citep{leonard2024}.\label{fig:jwst}}
\end{figure}

In Figure \ref{fig:jwst}A, we map the integrated band areas of the 3.5-$\mu$m H$_2$O$_2$ feature across Europa's leading hemisphere and show the corresponding map of Europa's $\nu$$_3$ CO$_2$ band, as obtained by \citet{TrumboBrown2023} (Figure \ref{fig:jwst}B). In agreement with the Keck observations \citep[][this work]{Trumbo2019h2o2}, we find that the strongest H$_2$O$_2$ absorption occurs in the leading-hemisphere chaos terrains Tara Regio ($\sim$85° W) and Powys Regio ($\sim$125° W) (Figure \ref{fig:jwst}A), which also exhibit the strongest CO$_2$ signatures (Figure \ref{fig:jwst}B). In fact, a simple linear fit between the H$_2$O$_2$ and CO$_2$ band strengths in the JWST data reveals a strong positive correlation (Pearson correlation coefficient of 0.94), as would be expected for the previously hypothesized effect of endogenic CO$_2$ on radiolytically produced H$_2$O$_2$ (Figure \ref{fig:jwst}C).

Two distinct peaks are resolved within the $\nu$$_3$ band of Europa's CO$_2$ (at 4.25 $\mu$m and at 4.27 $\mu$m) \citep{TrumboBrown2023}, along with a combination band at 2.7 $\mu$m \citep{Villanueva2023}. As these signatures might indicate different CO$_2$ host materials or trapping mechanisms, we also assess the correlation of H$_2$O$_2$ with the individual depths of each CO$_2$ peak within the $\nu$$_3$ band and with the strength of the 2.7-$\micron$ feature. The Pearson correlation coefficients (R) calculated with respect to the 4.25-$\mu$m peak, the 4.27-$\mu$m peak, and the 2.7-$\micron$ band---with coefficients of 0.92 (P\(<\)0.001), 0.92 (P\(<\)0.001), and 0.90 (P\(<\)0.001), respectively---are consistent with that calculated with respect to the overall $\nu$$_3$ band (R=0.94; P\(<\)0.001). As all three CO$_2$ signatures are individually the strongest within Tara Regio ($\sim$85° W) and Powys Regio ($\sim$125° W) \citep{TrumboBrown2023, Villanueva2023}, it is unsurprising that we cannot separate a correlation between H$_2$O$_2$ and a particular CO$_2$ phase from a simple strong correlation with the overall abundance of CO$_2$ regardless of its trapping mechanisms or host materials. Nevertheless, the strong pixel-to-pixel correspondence suggests that the presence of CO$_2$ in some form may be an important factor affecting the distribution of H$_2$O$_2$ in the chaos terrains, though the possibility remains that H$_2$O$_2$ could be associated with chaos terrains for an independent reason.

\section{Discussion} \label{sec:discussion}
Our findings suggest that Europa's H$_2$O$_2$ is largely insensitive to the daily temperature changes it experiences. We find no strong evidence to support the idea that H$_2$O$_2$ builds up efficiently overnight, and then becomes depleted across Europa's day, as hypothesized from laboratory work demonstrating an inverse dependence of equilibrium H$_2$O$_2$ concentrations on ice temperature \citep{HandCarlson2011}. We suggest that this apparent discrepancy might be due to the short timescales of Europa's temperature changes, \edit1{though we also note that the observed magnitude of the influence of temperature also varies widely across different irradiation experiments and may be part of the explanation as well \citep[e.g.,][]{GomisLeto2004, LoefflerEtAl2006, ZhengEtAl2006}}. Indeed, conditions directly analogous to Europa's diurnal cycles have, to our knowledge, not been simulated in the lab, and the timescale for the reaction of the dynamic equilibrium of production and destruction of H$_2$O$_2$ in impure ice to temperature perturbations is uncertain. \citet{HandCarlson2011} infer that, if starting from pristine ice (with no H$_2$O$_2$ or other non-ice materials, like CO$_2$) and irradiating at a constant surface temperature (as was done in their experiments), it would take an estimated $\sim$48 hours of irradiation on the trailing hemisphere and approximately two weeks on the leading hemisphere to reach equilibrium \citep{HandCarlson2011,Carlson2009}. These timescales are extremely short in geologic terms, suggesting that a dynamic equilibrium with respect to H$_2$O$_2$ is almost instantly achieved following any resurfacing events, but they do not directly enable precise predictions of diurnal variations. Future irradiation experiments that first reach equilibrium H$_2$O$_2$ concentrations and then continuously monitor changes in H$_2$O$_2$ absorption, while simulating the temperature fluctuations observed on Europa through controlled heating and cooling, could help further investigate the timescales needed for Europa's H$_2$O$_2$ to respond to local temperature changes. 

Though our IRTF and Keck results contradict the hypothesized temperature effects, our JWST results showing a strong correlation between  H$_2$O$_2$ and CO$_2$ within large-scale, leading-hemisphere chaos regions support the previously proposed hypothesis that CO$_2$ may be an important factor controlling the geographic distribution of H$_2$O$_2$ on Europa \citep{Trumbo2019h2o2}. \edit1{Certain laboratory experiments have shown that electron-accepting contaminants, such as O$_2$ and CO$_2$, have the potential to enhance H$_2$O$_2$ yields \citep{MooreHudson2000,Strazzulla2005}, potentially by inhibiting the breakdown of newly formed H$_2$O$_2$ by consuming the destructive electrons produced as the ice continues to be irradiated \citep{MooreHudson2000}}. Indeed, recent work continues to build on these prior experiments and appears to confirm the augmenting effects of small amounts of CO$_2$ more robustly \edit1{\citep{Mamo2023,Raut2024}}, which further suggests that the spatial correlation between Europa's CO$_2$ and H$_2$O$_2$ reflects this effect, rather than coincidence. 

Though CO$_2$ may explain the surprising association of radiolytically produced H$_2$O$_2$ with endogenic terrains on the leading hemisphere, it remains unclear how important it is for the trailing side. It is possible that the lower water-ice fraction \citep[e.g.,][]{BrownHand2013,Fischer2015, Ligier2016} and warmer temperatures \citep[e.g.,][]{Rathbun2010, Trumbo_2018} on the trailing side may help explain its comparative lack of H$_2$O$_2$. The potential chemical reaction with SO$_2$ \citep{LoefflerHudson2013} also remains a compelling possibility. As there are few existing constraints on the possibility of CO$_2$ on Europa's trailing hemisphere \citep{Carlson2009, HansenMcCord2008}, the magnitude of its influence relative to these other effects remains an open question. Future JWST observations of the trailing side would help further explore the H$_2$O$_2$-CO$_2$-chaos connection and in the context of the added complications of sulfur radiolysis \citep[e.g.,][]{Carlson2002, Carlson2009}.

H$_2$O$_2$ was recently discovered on Ganymede and was found to be associated with its cold, icy, and highly irradiated polar regions \citep{trumbo2023_h2o2Ganymede}. This discovery contrasts with the distribution of Europa's H$_2$O$_2$ in warm, ice-poor, large-scale chaos terrains at low-to-mid latitudes, and instead aligns better with laboratory expectations on the effects of water-ice availability, temperature, and irradiation \citep[e.g.,][]{ZhengEtAl2006,LoefflerEtAl2006,MooreHudson2000,HandCarlson2011}. Indeed, Ganymede's equatorial latitudes, where little H$_2$O$_2$ is observed, are shielded from irradiation by its intrinsic magnetic field \citep[e.g.,][]{khurana2007origin, Poppe2018} and are warmer than Europa's equator \citep[e.g.,][]{orton1996galileo, Spencer1999, dekleer2021ganymede}. Thus, though such factors do not seem to control the geographic distribution of H$_2$O$_2$ on Europa, they may still be important on Ganymede. Interestingly, however, Ganymede's surface also features ubiquitous CO$_2$ \citep[e.g.,][]{Hibbitts2003}, which JWST data suggest exists in multiple trapped phases across the surface \citep{BockeleeMorvan2024}. The CO$_2$ appears most abundant in the equatorial regions, where it is suggested to be trapped in dark and non-icy substrates \citep{Hibbitts2003, BockeleeMorvan2024}, which would most likely separate it from the production of H$_2$O$_2$ in water ice. However, in the polar regions, the CO$_2$ may be trapped within amorphous water ice and is spatially coincident with the H$_2$O$_2$  \citep{BockeleeMorvan2024}. This opens up the possibility of a similar influence on H$_2$O$_2$ as seems likely on Europa and is consistent with the possibility that CO$_2$, in addition to the low temperatures, more abundant water ice, and preferential radiolysis driven by Jovian magnetospheric particles, might help enhance H$_2$O$_2$ abundances at Ganymede's poles.

JWST observations of Callisto, in contrast, reveal no signs of H$_2$O$_2$ \citep{Cartwright_2024}, despite its widespread CO$_2$ \citep{Hibbitts2000,Hibbitts2002, Cartwright_2024}. The lack of H$_2$O$_2$ on Callisto might be related to multiple factors. First, Callisto's surface, although featuring localized bright and ice-rich knobs \citep{moore2004}, appears to be primarily characterized by a dark, ice-poor lag deposit \citep{moore1999mass, Cartwright_2024}, which would restrict the availability of water ice for radiolysis. Moreover, Callisto's distance from Jupiter results in a much weaker impact of charged-particle irradiation on its surface, and the flux of charged particles from Jupiter's magnetosphere is hundreds of times lower than at Europa \citep{COOPER2001, Johnson2004}. Additionally, Callisto (like Ganymede's equatorial latitudes) has a higher surface temperature (representative equatorial daytime temperature $\sim$155 K) than Europa (representative equatorial daytime temperature of $\sim$120 K)\citep{Spencer1999,Camarca_2023,Hanel1979, Rathbun2010, Trumbo_2018, Spencer1987}. Collectively, these factors may inhibit the production of H$_2$O$_2$ at Callisto to the point where the presence of CO$_2$ is not enough to result in detectable H$_2$O$_2$.

Looking at H$_2$O$_2$ across the Galilean system may thus help us understand the factors controlling its presence on each moon. CO$_2$ appears likely important for H$_2$O$_2$, on Europa and potentially also on Ganymede. Surface temperature, while not important for spatial or temporal variations of H$_2$O$_2$ on Europa, does align with the  influences expected from laboratory studies \citep[e.g.,][]{ZhengEtAl2006,LoefflerEtAl2006,MooreHudson2000,HandCarlson2011} when looking at the icy Galilean satellites as a whole. Future JWST observations of the icy satellites of Saturn and icy Kuiper Belt Objects---like Charon (Pluto's largest moon), where both CO$_2$ and H$_2$O$_2$ were recently detected \citep{Protopapa2023Unveiling}---may enhance our understanding of the complex interplay of factors influencing the radiolytic production of H$_2$O$_2$, thereby providing deeper insights into this important process in the outer solar system.

\section{Conclusion} \label{sec:conclusion}
In this study, we use a combination of IRTF, Keck, and JWST observations of Europa to asses the influence of temperature and CO$_2$ on Europa's H$_2$O$_2$. We find at most minimal changes in H$_2$O$_2$ as a result of local temperature variations on timescales associated with Europa's eclipse by Jupiter and diurnal cycle, suggesting a surprising temperature insensitivity that appears at odds with some laboratory studies. We suggest that this apparent discrepancy might be explained if the diurnal timescales are short compared to those needed for the H$_2$O$_2$ concentrations to respond to a temperature perturbation and recommend laboratory experiments to explore this further. Using JWST NIRSpec data, we find a clear and strong correlation between Europa's H$_2$O$_2$ and CO$_2$, supporting the previously suggested hypothesis that endogenic CO$_2$ may enhance H$_2$O$_2$ abundances and thereby explain the unexpected concentration of H$_2$O$_2$ in low-latitude chaos terrain. This correlation is also in agreement with limited past and recent laboratory work suggesting that CO$_2$ can act as an electron-scavenging contaminant that increases H$_2$O$_2$ yields by inhibiting its breakdown. Future spatially resolved observations on Europa's trailing hemisphere can further test this correlation and explore the possible influence of sulfur-bearing, trailing-hemisphere components as well.

\acknowledgments
The IRTF/Spex data presented were obtained at the Infrared Telescope Facility, which is operated by the University of Hawaii under contract 80HQTR19D0030 with the National Aeronautics and Space Administration. The Keck AO data presented were obtained at the W. M. Keck Observatory, which is operated as a scientific partnership among the California Institute of Technology, the University of California, and the National Aeronautics and Space Administration. The Observatory was made possible by the generous financial support of the W. M. Keck Foundation. The authors wish to recognize and acknowledge the very significant cultural role and reverence that the summit of Maunakea has always had within the indigenous Hawaiian community. We are most fortunate to have the opportunity to conduct observations from this mountain. The JWST data were obtained from the Mikulski Archive for Space Telescopes (MAST) at the Space Telescope Science Institute, which is operated by the Association of Universities for Research in Astronomy under NASA contract NAS 5-03127 for JWST. These observations are associated with program \#1250. The authors acknowledge H. Hammel and the GTO team led by PI G. Villanueva for developing their observing program with a zero-exclusive-access period. The specific observations analyzed can be accessed via \dataset[DOI: 10.17909/e269-sm44]{https://doi.org/10.17909/e269-sm44}. We thank Ryleigh Davis for fruitful discussions and Jonathan Lunine for insightful comments and constructive suggestions. We thank Erin Leonard for providing shape files for Europa's geologic units.

\vspace{5mm}
\facilities{IRTF/Spex, Keck AO, JWST}
\software{Astropy \citep{astropy2013,astropy2018,astropy2022}, Scikit-image \citep{scikitimage2014}, Scipy \citep{SciPy2020}, Cartopy \citep{Cartopy2015}, Astroquery \citep{astroquery2019}, Spextool \citep{Cushing2004}}

\clearpage
\appendix
\begin{figure*}[ht!]
\figurenum{A1}
\centering
\includegraphics[scale=0.75] {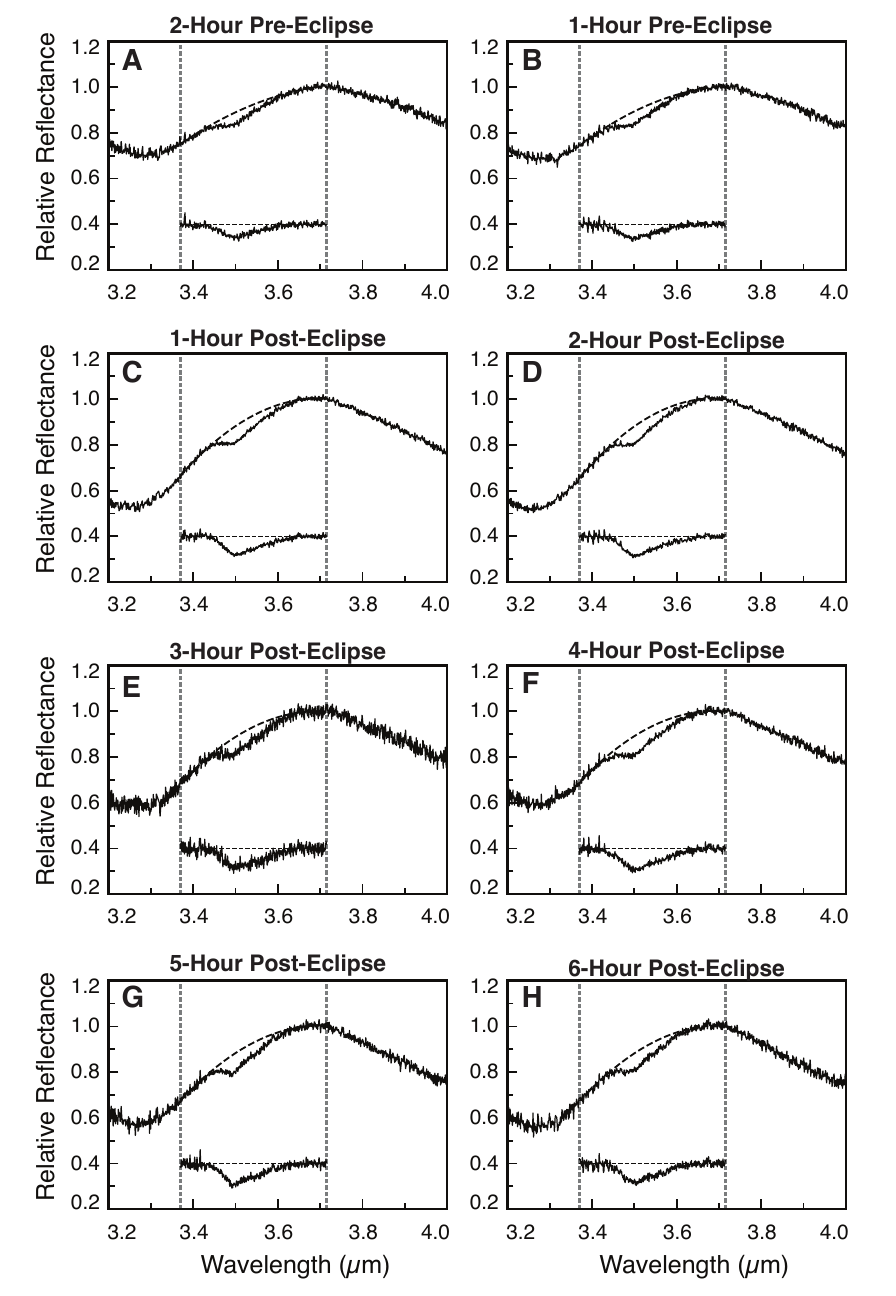}
\caption{Averaged spectra for each $\sim$1-hour time window of [A] 2 hours pre-eclipse, [B] 1 hour pre-eclipse, [C] 1 hour post-eclipse, [D] 2 hours post-eclipse, [E] 3 hours post-eclipse, [F] 4 hours post-eclipse, [G] 5 hours post-eclipse, and [H] 6 hours post-eclipse. All spectra are normalized to their individual peaks in the 3.6–3.7 $\micron$ region. Dashed gray lines outline the H$_2$O$_2$ band. Second-order polynomial continua are indicated by the black dashed curves. Continuum-removed absorptions are included. We note that the 3-hour post-eclipse spectrum is noisier due to a shorter observation time (necessitated by weather conditions), as indicated in Table \ref{table:obs}. A factor of change $<$2 is shown in the H$_2$O$_2$ band throughout the 8-hour time period surrounding eclipse. The slight changes to the background water-ice absorption bands suggest likely rotational effects as Europa rotates from a more trailing/sub-Jovian geometry towards a more sub-Jovian/leading geometry. More discussion is included in section \ref{sec:results1}.}
\label{fig:supp_irtf}
\end{figure*}
\newpage

\begin{figure*}[ht!]
\figurenum{A2}
\plotone{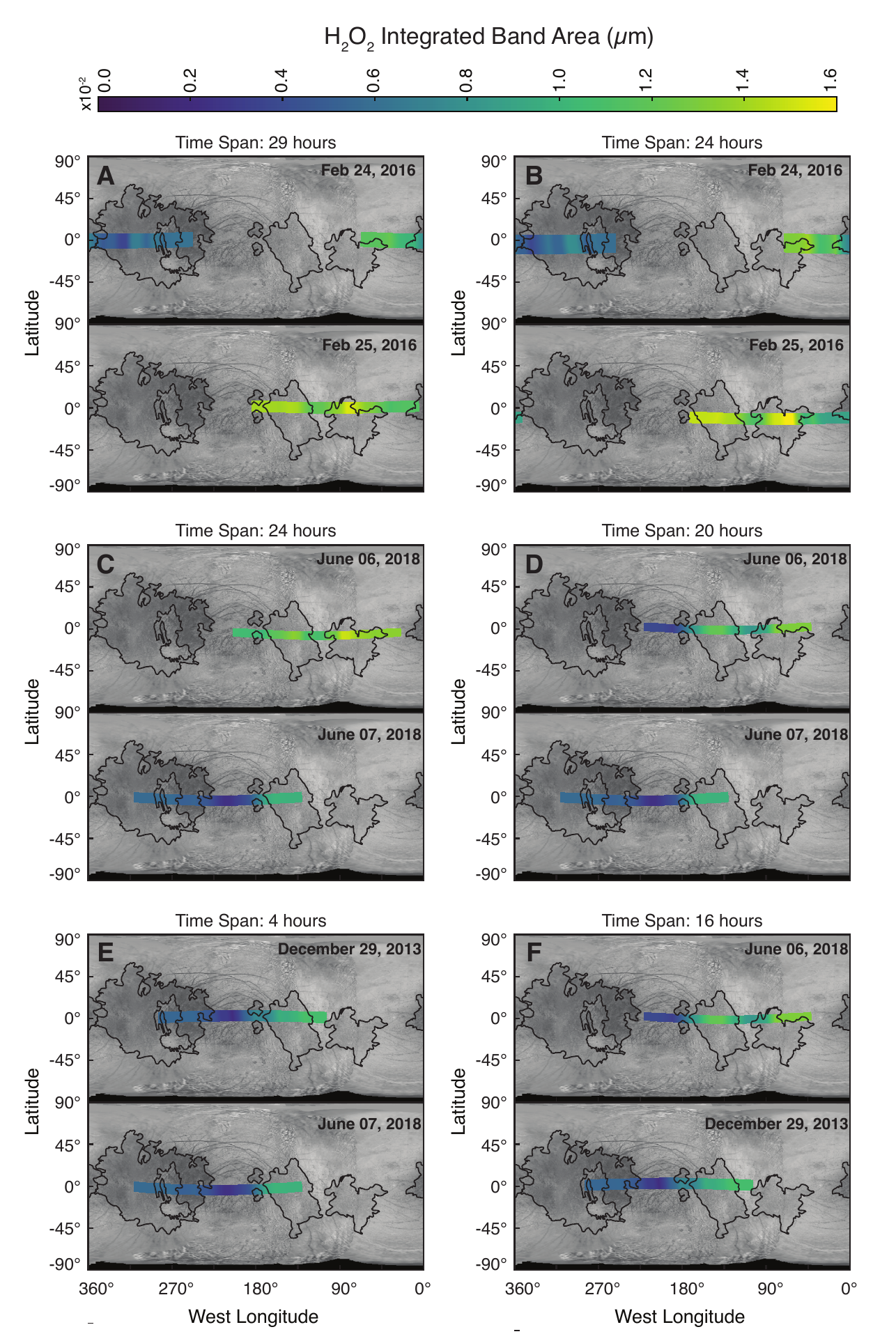}
\label{fig:supp}
\end{figure*}

\newpage
\begin{figure} [ht!]
\figurenum{A2}
  \caption{Additional representative paired slits to illustrate the lack of a strong diurnal trend.  We examine a total of 5 Keck slit pairs observed on adjacent nights in 2016 and 2018, one of which is shown in Figure \ref{fig:keck}, and 2 slit pairs observed across different years. Time spans for each paired comparison are labeled on top. [A] Paired slits from consecutive nights obtained on February 24, 2016 [top] and February 25, 2016 [bottom], when overlapped regions ($\sim$18-36\degree{}W) are in the morning and afternoon, respectively. [B] Paired slits slightly south of A with a slightly shorter time span, obtained on February 24, 2016 [top] and February 25, 2016 [bottom], when overlapped regions ($\sim$20-45\degree{}W) are in the morning and afternoon, respectively. [C] Paired slits obtained from consecutive nights on June 06, 2018 [top] and June 07, 2018 [bottom], when overlapped regions ($\sim$155-179\degree{}W) are in the morning and afternoon, respectively. [D] Paired slits from consecutive nights obtained on June 06, 2018 [top] and June 07, 2018 [bottom], when overlapped regions ($\sim$157-180\degree{}W) are in the morning and afternoon, respectively. [E] Paired slits obtained across years on December 29, 2013 [top] and June 07, 2018 [bottom], when overlapped regions ($\sim$155-182\degree{}W) are in the morning and afternoon, respectively. [F] Paired slits obtained across years on June 06, 2018 [top] and December 29, 2013 [bottom], when overlapped regions ($\sim$155-181\degree{}W) are in the morning and afternoon, respectively. For all pairs and overlapping regions examined, the measured band area changes from morning to afternoon are within -32$\%$ to 44$\%$ (factor of 0.68-1.44). These results suggest the absence of a strong diurnal trend and a lack of sensitivity in H$_2$O$_2$ to temperature variations over the assessed time scales. The slits also present a clear difference between low-latitude chaos and low-latitude plains terrain in east/west slits crossing the leading chaos regions [A-D] and a depletion of H$_2$O$_2$ west of the anti-Jovian point around 210°W, corresponding to a gap between the large-scale chaos regions of the leading and trailing hemispheres [C-F]. Leading and trailing chaos regions are outlined in black \citep{leonard2024}. Background basemap credit: \citet{USGSbasemap}.}
\end{figure}

\begin{figure*}[ht!]
\figurenum{A3}
\epsscale{0.9}
\plotone{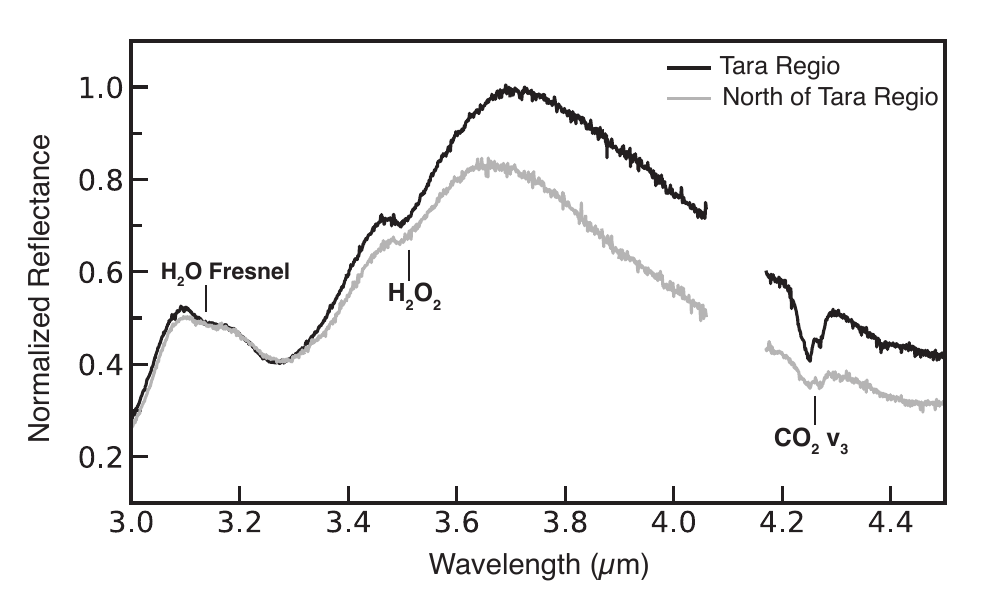}
\caption{Representive JWST spectra used for Figure \ref{fig:jwst}, showing both the H$_2$O$_2$ and the CO$_2$ bands in the Cycle 1 JWST data recently used to map Europa's leading-hemisphere CO$_2$ \citep{TrumboBrown2023,Villanueva2023}. The black spectrum is extracted from a pixel within Tara Regio (92\degree{}W, 17\degree{}S) and the gray spectrum is extracted from a pixel north of Tara Regio (57\degree{}W, 19\degree{}N). A clear enrichment of both H$_2$O$_2$ and CO$_2$ is shown in the Tara Regio spectrum.}
\label{fig:supp_JWST}
\end{figure*}

\clearpage
\bibliography{GalRefs}{}
\bibliographystyle{aasjournal}

\end{document}